\documentclass[revsymb,aps,prb,10pt]{revtex4}

\begin{document}
\title{Quantum dissipation theory of slow magnetic relaxation mediated by
domain-wall motion in one-dimensional chain compound [Mn(hfac)$_{\text{2}}$BNO$_{\text{H}%
} $].}
\author{A.S. Ovchinnikov\cite{byline}, I.G. Bostrem, V.E. Sinitsyn, A.S.
Boyarchenkov, N.V. Baranov}
\address{Department of Physics, Ural State University, 620083, Ekaterinburg, Russia}
\author{K. Inoue}
\address{Department of Chemistry, Faculty of  Science, Hiroshima University, Higashi-Hiroshima, 739-8526, Japan}

\today 

\begin{abstract}
Based on a quantum dissipation theory of open systems, we present a
theoretical study of slow dynamics of magnetization for the ordered state of
the new molecule-based magnetic complex [Mn(hfac)$_{\text{2}}$BNO$_{\text{H}%
} $] composed from antiferromagnetically coupled ferrimagnetic $(5/2,1)$
spin chains. Experimental investigations of the magnetization process in
pulsed fields have shown that this compound exhibits a metamagnetic AF-FI
transition at a critical field in the order of the interchain coupling. A strong
frequency dependence for the ac-susceptibility has been revealed in the
vicinity of the AF-FI transition and was associated with an AF-FI interface
kink motion. We model these processes by a field-driven domain-wall motion
along the field-unfavorable chains correlated with a dissipation effect due
to a magnetic system-bath coupling. The calculated longitudinal
magnetization has a two-step relaxation after the field is switched off and
are found in good agreement with the experiment. The relaxation time
determined from the imaginary part of the model ac-susceptibility agrees
qualitatively with that found from the remanent magnetization data.
\end{abstract}

\pacs{PACS numbers: 72.25.Rb,61.20.Lc,67.40.Fd}
\maketitle

\section{Introduction}

Magnetic resonance effects due to switching of spins by a time-dependent
magnetic field and relaxation measurements are widely used to study magnetic
materials. They may give information about the main mechanism of the
magnetization change in samples. When a magnetic field varies, magnetic
materials which exhibit a hysteresis present a characteristic time
dependence of the magnetization due to the multiplicity of available
metastable states. In many relaxing magnetic systems the time dependence can
be described by $M=M_0-S\ln t$, where $S$ is the magnetic viscosity of the
system. The viscosity is a consequence of thermal activation of irreversible
domain processes such as the domain wall motion and the nucleation of
domains of the reverse magnetization. The logarithmic change with time $%
M\sim \ln t$ is predicted if there is a distribution of energy barriers or
time-dependent activation energies present in the material. A simple Debye
relaxation $M=M_0\exp (-t/\tau )$ arises from a single-barrier activation
mechanism \cite{StrWoo49,Gau86}. The barriers associated with the relaxation
process are of two types. The intrinsic barriers arising, for example, from
the magnetic anisotropy contribute to the reversal of magnetization, whereas
the barriers due to the pinning of domain walls are generally attributed to
the defects in the materials. Both types of barriers are responsible for the
pronounced metastability (hysteresis phenomena) of magnetic systems.

Recent efforts in synthetic chemistry provide a number of low-dimensional
magnetic systems that show the slow relaxation of the magnetization, for
example, this effect was found in one-dimensional ($1D$) anisotropic
ferrimagnetic chains named as single chain magnets (SCM) \cite{Novak05}. The
slow magnetic relaxation in the paramagnetic phase has been observed with ac
susceptibility and SQUID magnetometry measurements in the real quasi-$1D$
ferrimagnetic compound [Co(hfac)$_2$NITPhOMe] \cite{Can2001}. An Arrhenius
behavior with the activation energy $\Delta \sim $ 152 K, which is of order
of the intrachain exchange interaction between alternating Co$^{2+}$ and
organic NITPhOMe spins, has been observed for 10 decades of relaxation time
and found to be consistent with the Glauber model \cite{Glaub63}. The
relaxation was also studied by monitoring the decay of the longitudinal
magnetization, which was found to be exponential. The slow relaxation is
governed by the uniaxial anisotropy seen by each spin on the chain and
magnetic correlations between the spins.

Recently, the ac susceptibility and magnetization in steady and pulsed
fields have been measured for the new molecule-based magnetic complexes
[Mn(hfac)$_{\text{2}}$BNO$_{\text{R}}$] (R=H,Cl) with ferromagnetically (Cl)
or antiferromagnetically (H) ordered ferrimagnetic chains composed of S=1
(biradical) and S=5/2 (Mn$^{2+}$) \cite{Bar2003}. It has been found that the
change in magnetization in these compounds under application of a magnetic
field below the three-dimensional (3D) ordering temperature (T$_{\text{N }}$%
=5.5 K (H) and T$_{\text{C }}$=4.8 K (Cl)) is a slow dynamical process which
presumably originates from their strong one-dimensional character, i.e.
because of the weakness of the interchain exchange ($J$) interaction in
comparison with intrachain ($J^{\prime }$) one ($J/J^{\prime }\sim 10^{-3}$%
). The thermally activated change of the remanent magnetization in [Mn(hfac)$%
_{\text{2}}$BNO$_{\text{H}}$] after switching off the field looks quite
different from that in [Mn(hfac)$_{\text{2}}$BNO$_{\text{Cl}}$]. Figure 1
displays time dependence of the remanent magnetization at 1.5 K for these
compounds for comparison. The large value of the relaxation times of both
processes and their existence well below $T_{3D}$ lead to the suggestion
that these relaxation processes are related to the development and motion of
magnetic domains either with a wide distribution of energy barriers (R=Cl)
or a single energy barrier height (R=H).

Indeed, the change of the magnetization of [Mn(hfac)$_{\text{2}}$BNO$_{\text{%
H}}$] with time during and after application of a pulsed field is controlled
by the direct AF-FI and inverse FI-AF transitions. For the applied pulse
with a duration of 20 ms and amplitude of 4 T, the direct AF-FI transition
is completed within a time of less than 2 ms. The high rate of the direct
transition is due to the high value of the magnetic field in comparison with
the critical field of the AF-FI transition (0.03 T in steady fields and
about 0.2 T in pulsed fields). The large relaxation time ($\sim $500 ms)
after removal of the pulsed field implies very slow dynamics of the
magnetization during the inverse transition from the field-induced FI state
to the initial AF state in zero field. The slow dynamics of the
magnetization in this material was suggested to be be controlled by the
domain wall motion along the separate field-unfavorable chains \cite{Bar2003}. 
The energy barrier which hampers the reversal of the magnetization is
originated by the antiferromagnetic interchain coupling that is reminiscent
of the molecular clusters or single cluster magnet (SCM) where the energy
barrier is due to the magnetic anisotropy.

The fast relaxation during the field change and the slow relaxation due to
the transitions between the different metastable states involves the release
of the magnetic energy which cause local heating. The process of energy
losses in [Mn(hfac)$_{\text{2}}$BNO$_{\text{H}}$] is likely due to the
motion of domain boundaries within the chains and is connected with the
coercetivity losses and the radiation of phonons. Finally, the energy of the
magnetic system is transformed into the phonon energy, that is, into the
heat. Due to the magnetic system-bath coupling we may explain the
experiments under the condition of isothermal relaxation, and, as a
consequence, in an assumption that self-heating in the magnetic relaxation
does not destroy 3D order.

\section{Preliminaries}

The choice of a proper formalism of the quantum dissipation theory depends
essentially on the real physical media and real process of the energy
transfer. The interaction between a quantum system and its environment is
the physical factor responsible for the relaxation process in the system.
Such a relaxation process can be provided by the environment if it acts as a
bath. This arises the fundamental problem of description of the relaxation
dynamics for a system simultaneously interacting with a heat bath and a
time-dependent driving field.

The question now arises as to what is a nature of heat bath. We note that
the spin system in the external dc bias field in ac-susceptibility
measurements or in the pulsed field in magnetization measurements consists
of two weakly interacting parts, i.e. field-favorable ($S_1$) and
field-unfavorable ($S_2$) chains. Then, two distinct models for the bath are
assumed. The first one identifies the bath with a crystal lattice (phonon
bath). In this case, the energy levels of the bath are populated according
to the conventional Gibbs distribution. When a rapidly varying strong pulsed
field causes the AF-FI transition the spin temperature of the $S_2$ part
becomes much higher than the lattice and the $S_1$ part temperatures
(saturation process). After the field is off the whole spin quantum system
is driven to the thermal equilibrium characterized by Gibbs's density matrix
owing to energy exchange between the $S_1$, $S_2$ systems and the heat bath.
Following the fundamental Bogoliubov's procedure of the contraction of
description with the accompanying hierarchy of relaxation times \cite{Bogo70}
we may conclude that the weak interchain interactions are related to
long-time relaxation process with the characteristic time equals to the
characteristic time scale of the experiment $1-10^3$ ms. It is assumed that
another interactions, including usual spin-phonon ones, are strong enough
and have associated correlation effects with shorter relaxation times, i.e.
much smaller than the duration of the experiment. We should eliminate them
as an irrelevant information to characterize the macrostate. Apparently, a
possible approach to this irreversible process may be performed within the
nonequilibrium statistical operator (NSO) method largely elaborated by
Zubarev and co-workers \cite{Zuba69,Zuba97}, which is a large generalization
of the Gibbs's theory. The results of the NSO analyses will be given by us
elsewhere.

The second model of the bath is based on the suggestion that the relaxation
process is related with the development and motion of magnetic domains in
the field-unfavorable chains $S_2$ in a driving field. The recent study of
the behavior of domain walls in Ising ferromagnetic chains yields the
following picture \cite{HaRu99}. In zero field the interface between two
domains of oppositely oriented spins, a kink, moves left or right with the
probability $p=1/2$, which may be interpreted as a random walk. Two such
kinks can meet via diffusive motion. Once there is only one spin left
between two kinks they annihilate with probability $p\sim 1$ in the next
time step and the two domains merge. Switching on the external driving field
causes domains with spins parallel to the field to start growing. The kinks at
the end of such a cluster move outwards one step during each time step where
the field remains favorable. If the field switches into unfavorable
direction, the cluster shrinks again (breathing behavior). In this process
the small domains and the kinks associated with them will be eliminated
rapidly. In the Ising model a domain wall width is simply the lattice
constant. On the contrary, the domain walls in the Heisenberg chains are
much wider due to the strong exchange interactions involved. The closest
analog of such wall is the soliton in a magnetic chain \cite{Jongh82}. 
Therefore another model, appropriate for this domain wall
relaxation dynamics, may be applied for the bath. In this model a quantum
system (the field-unfavorable chain with a kink) interacts weakly with the
environment, i.e. with the nearest surrounding field-favorable chains and
the lattice (Fig. 2). The bath is again considered as a quantum statistical
system being at equilibrium. The domain walls oscillate around their
equilibrium position under a varying ac field (the process far from a
saturation) or move reversibly under a time-dependent pulsed field (the
saturation process). The bath provides an existence of random fields
(Langevin forces) created by the environment which interacts with variables
of the quantum system (Langevin dissipative modes). Due to the system-bath
interactions the domain wall dynamics becomes irreversible and the system
relaxes from its initial nonequilibrium state to the equilibrium one when
the field is off.

A quantum dissipative theory (QDT) with the system-bath interaction being
treated rigorously at the second-order cumulant level for reduced dynamics
has been recently constructed for open quantum systems \cite{XuYan02}. The
theory belongs to a class of widely used quantum master equations, such as
the Bloch-Redfield theory \cite{Bloch53,Red57} and a class of Fokker-Plank
(FP) equations\cite{Dek81,Zuba97}, and is valid for arbitrary bath
correlation functions and time-dependent external driving fields. The QDT-FP
formulation constitutes a theoretical framework for the present study of
dissipative processes in the molecule-based magnetic complexes [Mn(hfac)$_{%
\text{2}}$BNO$_{\text{H}}$].

\section{Quantum dissipation theory}

The key theoretical quantity in a quantum dissipation is the reduced density
operator $\rho (t)\equiv Tr_B\rho _T(t)$, i.e. the partial trace of the
total system and bath composite $\rho _T(t)$ over all the bath degrees of
freedom. For a system dynamical variable $A$, its expectation value 
\[
\left\langle A(t)\right\rangle =Tr\left[ A\rho _T(t)\right] =Tr\left[ A\rho
(t)\right] 
\]
can be evaluated with the substantially reduced system degrees of freedom.
Quantum dissipation theory governs the evolution of the reduced density
operator $\rho (t)$, where the effects of bath are treated in a quantum
statistical manner.

The total composite Hamiltonian in the presence of classical external field
can be written as 
\begin{equation}
H_T=H(t)+H_B-\sum\limits_aQ_aF_a  \label{HT}
\end{equation}

Here $H(t)$ is the deterministic Hamiltonian that governs the coherent
motion of the reduced system density matrix and involves interaction with an
arbitrary external classical field $h(t)$. The system is embedded in a
dissipative bath ($H_B$) and the last term in Eq.(\ref{HT}) describes the
system-bath couplings, in which $\left\{ Q_a\right\} $ are Hermite operators
of the primary system and can be called the generalized dissipative modes.
The generalized Langevin forces 
\[
F_a(t)=e^{iH_Bt}F_ae^{-iH_Bt} 
\]
are Hermite bath operators in the stochastic bath subspace assuming Gaussian
statistics. Without loss of generality, their stochastic mean values are set
to $\left\langle F_a(t)\right\rangle _B=0$, where $\left\langle \ldots
\right\rangle _B$ denote the ensemble average over the initially stationary
bath density matrix $\hat{\rho}_B(0)$. The effects of Langevin forces on the
reduced primary system are therefore completely characterized by their
correlation functions $\tilde{C}_{ab}(t)=\left\langle
F_a(t)F_b(0)\right\rangle _B$. They satisfy the boundary conditions $\tilde{C%
}_{ab}(\pm \infty )=0$, and the detailed-balance and the symmetry relations $%
\tilde{C}_{ab}^{*}(t)=\tilde{C}_{ab}(t-i\beta )=\tilde{C}_{ba}(-t),$ where $%
\beta =1/kT$. This admits the Meier-Tannor parametrization $\tilde{C}%
_{ab}(t) $ in terms of exponential functions \cite{MeTan99} 
\[
\tilde{C}_{ab}(t\geq 0)\equiv \sum_m\nu _m^{ab}e^{-\zeta _m^{ab}t} 
\]
with the adjustable parameters $\nu _m^{ab}$, $\zeta _m^{ab}$. These
parameters are in general complex but, for simplicity, we take $\zeta
_m^{ab} $ to be real and positive.

The frequency-domain symmetry relation reads as $C_{ab}^{*}(\omega
)=C_{ba}(\omega )$, and the detailed-balance relation in terms of spectral
functions is $C_{ba}(-\omega )=e^{\beta \omega }C_{ab}(\omega )$. Using the
generalized bath interaction spectral density function $J_{ab}(\omega
)=C_{ba}(-\omega )-C_{ab}(\omega )$ obeying the symmetry relations $%
J_{ab}(\omega )=-J_{ba}(-\omega )=J_{ba}^{*}(\omega )$ we have $%
C_{ab}(\omega )=J_{ab}(\omega )/\left( e^{\beta \omega }-1\right) $.

We will use the reduced Liouville equation, i.e. the equation of motion for
the reduced density matrix $\rho (t)$, in the partial ordering prescription 
\cite{Yan98} 
\begin{equation}
\dot{\rho}(t)=-iL(t)\rho (t)-R(t)\rho (t),  \label{RLE1}
\end{equation}
which is characterized by the local-time kernel $R(t)$. The Liouvillian $L$
is the commutator of the reduced system Hamiltonian $H(t)$ in the presence
of external classical field 
\begin{equation}
L(t)\hat{A}\equiv \left[ H(t),\hat{A}\right] ,  \label{Liov}
\end{equation}
and the superoperator $R(t)$ can be formulated in terms of the system-bath
interaction. In the standard approximation of the weak-coupling limit in
which the system-bath interaction is considered only up to second order the
dissipation term is 
\begin{equation}
R(t)\rho (t)\equiv \sum\limits_a\left[ Q_a,\tilde{Q}_a(t)\rho (t)-\rho (t)%
\tilde{Q}_a^{\dagger }(t)\right] ,  \label{DIS}
\end{equation}
where $\tilde{Q}_a(t)$ is the non-Hermitian relaxation operator in the
Hilbert space 
\begin{equation}
\tilde{Q}_a(t)=\sum\limits_b\int\limits_{-\infty }^td\tau \;\tilde{C}%
_{ab}(t-\tau )G(t,\tau )Q_b.  \label{Qcmn}
\end{equation}
The Liouville-space propagator $G(t,\tau )$ associated with $L(t)$ is
defined via 
\[
i\frac \partial {\partial t}G(t,\tau )=L(t)G(t,\tau ). 
\]
It can be defined in terms of the Hilbert-space Green's function $\tilde{G}%
(t,\tau )$ via the relation for an arbitrary operator $\hat{A}$ 
\[
G(t,\tau )\hat{A}\equiv \tilde{G}(t,\tau )\hat{A}\tilde{G}^{\dagger }(t,\tau
), 
\]
where we treat $\hat{A}$ in the left-hand side as a vector in Liouville
space.

The reduced system Hamiltonian in the presence of external classical field
can be written as 
\[
H(t)=H_s+H_{sf}(t), 
\]
where $H_s$ is the time-independent, field-free Hamiltonian, whereas $%
H_{sf}(t)$ is the interaction between the system and the external classical
field $h(t)$. We further define similarly to Eq.(\ref{Liov}) the Liouville
superoperators $L_s$ and $L_{sf}(t)$ corresponding to the reference
Hamiltonians. The identity 
\begin{equation}
G(t,\tau )=G_s(t,\tau )-i\int\limits_\tau ^td\tau ^{\prime }\;G(t,\tau
^{\prime })L_{sf}(\tau ^{\prime })G_s(\tau ^{\prime },\tau )  \label{Iden1}
\end{equation}
obtained from the definitions 
\[
G(t,\tau )\equiv \hat{T}\exp \left( -i\int\limits_\tau ^t\left[
L_s+L_{sf}(\tau ^{\prime })\right] d\tau ^{\prime }\right) , 
\]
where the symbol $\hat{T}$ implies a time ordering, and 
\[
G_s(t,\tau )=G_s(t-\tau )=e^{-iL_s(t-\tau )} 
\]
allows us to separate the dissipation effects in Eq.(\ref{Qcmn}) into the
field-free part and the correlated driving-dissipation part. This yields 
\begin{equation}
\tilde{Q}_a(t)=\tilde{Q}_a^s-i\sum\limits_b\int\limits_{-\infty }^td\tau
\;\int\limits_\tau ^td\tau ^{\prime }\;\tilde{C}_{ab}(t-\tau )G(t,\tau
^{\prime })L_{sf}(\tau ^{\prime })G_s(\tau ^{\prime }-\tau )Q_b,  \label{Qt}
\end{equation}
where the field-free contribution $\tilde{Q}_a^s$ is time-independent and
given explicitly by 
\begin{equation}
\tilde{Q}_a^s=\sum\limits_b\int\limits_{-\infty }^td\tau \;\tilde{C}%
_{ab}(t-\tau )G_s(t-\tau )Q_b=\sum\limits_b\int\limits_0^\infty d\tau \;%
\tilde{C}_{ab}(\tau )e^{-iL_S\tau }Q_b=\sum\limits_bC_{ab}(L_S)Q_b.
\label{Qs}
\end{equation}
The equation (\ref{RLE1}) together with Eqs.(\ref{DIS},\ref{Qt},\ref{Qs})
provide a prescription for obtaining the reduced density operator up to the
second order in system-bath interaction. The underlining assumption is that 
the system-bath coupling is not strong enough, which makes the second order 
cumulant expression reasonable.   It is known that this approximation applies 
well to most dissipative systems in quantum optics, and to transport 
in mesoscopic systems \cite{LiLuo05}. In the strong system-bath regime a special 
technique is required, which goes beyond the second order approximation \cite{XueCui05,Schoel94}.

One of the traditional approaches to treat the problem is based on the associated quantum 
master equations. It focuses on the relation among various matrix elements of the
density operator in the time-independent $H$-eigenstate representation and is well 
suitable for the finite systems. For larger systems (spin chains, for example) the number of 
many-particle states increases dramatically and we cannot generally to solve all 
the microscopic equations. However, we can describe the system by macroscopic variables 
(domain wall position,  total magnetization of a chain) which fluctuate in a stochastic way. 
The Fokker-Plank (FP) equation arises as an equation of motion for the distribution function 
of the fluctuating macroscopic variables \cite{Risken}. The Eqs.(\ref{RLE1},\ref{DIS},\ref{Qt},\ref{Qs}) 
will serve as a starting formulations for deriving FP equations 
for observables of the reference system.

\section{Process far from a saturation: dynamic susceptibility}

Magnetic systems exhibiting relaxation phenomena can be characterized by the
complex ac susceptibility, $\chi (\omega )=\chi ^{\prime }-i\chi ^{^{\prime
\prime }}$, where the dispersion $\chi ^{\prime }$ and the absorption $\chi
^{^{\prime \prime }}$ are frequently dependent. Before moving on to the
technical details of the calculation, we mention briefly some of the
experimental results in the ac susceptibility measurements for the [Mn(hfac)$_{%
\text{2}}$BNO$_{\text{H}}$] compound that support the domain-wall motion
picture.

A strong frequency dependence both for $\chi ^{\prime }(\omega )$ and $\chi
^{^{\prime \prime }}(\omega )$ has been revealed in the bias dc fields of
0.025-0.03 T. The magnetization process in the field range 0.02-0.05 T is
accompanied by a remarkable hysteresis (of about 0.012 T), and, in addition,
a small remanent magnetization was detected after removal of the field.
These features are indicative of a magnetic phase transition of the first
order, which occurs through a mixed phase state, from antiferromagnetic
ordering of the chain magnetic moments to their parallel alignment in the
field-induced state \cite{Snoska}. In the region of the metamagnetic
transition, where the AF and FI\ phase coexist, the amplitude of a maximum
of both $\chi ^{\prime }$ and $\chi ^{^{\prime \prime }}$ decreases
significantly with increasing frequency, especially in the frequency region
from 1 to 50 Hz. From the large values of $\chi (\omega )$ for $\omega \sim
1 $ Hz in the vicinity of the AF-FI transition, we may suggest that
excitation of domain wall motion by a small oscillating field occurs more
effectively at low frequencies.

The complex magnetic ac susceptibility can be explained within approach,
where the magnetization $M$ is controlled by the field-induced sideways
motion of domain walls. In this case, the contribution of one domain wall to
the susceptibility is 
\[
\chi =\frac{\partial M}{\partial h}=\frac{\partial M}{\partial x}\frac{%
\partial x}{\partial h}, 
\]
and taking approximately $M\propto x$ as the magnetization increases due to
a wall displacement along the $x$ axis, one finds $\chi \propto \partial
x/\partial h$ \cite{Jongh82}.

A periodic domain wall motion caused by the external ac-field $h(t)=h_0\cos
\omega t$ is modeled by a well studied system, a driven Brownian oscillator
(DBO), with the Hamiltonian 
\[
H_s=\Omega \left( a^{\dagger }a+\frac 12\right) , 
\]
where the number of oscillator excitations $a^{\dagger }a$ corresponds to an
instant magnetization. The dissipation coupling mode $\hat{Q}$ 
\[
\hat{Q}=\frac 1{\sqrt{2}}\left( q_{+}\,\hat{a}^{\dagger }+q_{-}\,\hat{a}%
\right) 
\]
interacts with a stochastic bath force. Here, $q_{+}=q_{-}^{*}$ are complex
numbers, and $\hat{a}\,(\hat{a}^{\dagger })$ are the annihilation (creation)
operators of the oscillator with the frequency $\Omega $ determined by the
interchain coupling in an applied bias field, i.e. in the vicinity of the
AF-FI\ transition, and we hold only a single Langevin mode $Q$ in study.  
The operator $\hat{\mu}=\frac 1{\sqrt{2}}\left( \hat{a}^{\dagger }+\hat{a}%
\right) $ interacts with the ac-field and describes periodic domain wall
movement caused by $h(t)$. After introducing these definitions, Eq.(\ref{Qt}%
) can be transformed as follows 
\begin{equation}
\tilde{Q}(t)=\tilde{Q}^s-\frac i2\int\limits_0^\infty d\tau \;\lambda (\tau
)h\left( t-\tau \right) ,  \label{Lang1}
\end{equation}
where the system-bath coupling response 
\[
\lambda (\tau )=\int\limits_\tau ^\infty d\tau ^{\prime }\,\tilde{C}(\tau
^{\prime })\left[ q_{+}\,e^{i\Omega (\tau ^{\prime }-\tau
)}-q_{-}\,e^{-i\Omega (\tau ^{\prime }-\tau )}\right] 
\]
is given explicitly as 
\begin{equation}
\lambda (\tau )=\frac{q_{+}\nu _m}{\zeta _m-i\Omega }e^{-\zeta _m\tau }-%
\frac{q_{-}\nu _m}{\zeta _m+i\Omega }e^{-\zeta _m\tau }  \label{chit}
\end{equation}
with the aid of Meier-Tannor parametrization. As usual, a summation is to be
made for the repeated index $m$. In the following calculation we choose $%
q_{\pm }=1$ for simplicity, i.e. the dissipation is the same both for left
and right domain-wall displacements.

Substituting (\ref{chit}) into Eq.(\ref{Lang1}), followed by some minor
algebra, we get 
\[
\tilde{Q}(t)=\tilde{Q}^s+\frac{h_0\Omega \,\nu _m}{\left( \zeta _m^2+\Omega
^2\right) \left( \zeta _m^2+\omega ^2\right) }\left( \zeta _m\cos \omega
t+\omega \sin \omega t\right) . 
\]
The time-local dissipation superoperator is 
\[
R(t)\rho (t)\equiv \left[ Q,\tilde{Q}(t)\rho (t)-\rho (t)\tilde{Q}^{\dagger
}(t)\right] =R_s\rho (t)+i\xi (t)\left[ Q,\rho (t)\right] , 
\]
where $R_s$ is the field-free dissipation 
\[
R_s\rho (t)\equiv \left[ Q,\tilde{Q}^s\rho (t)-\rho (t)\left( \tilde{Q}%
^s\right) ^{\dagger }\right] . 
\]
The effective local-field correction, acting on the system via $Q$ is 
\[
\xi (t)=\frac{2h_0\Omega \,\nu _m^{^{\prime \prime }}}{\left( \zeta
_m^2+\Omega ^2\right) \left( \zeta _m^2+\omega ^2\right) }\left( \zeta
_m\cos \omega t+\omega \sin \omega t\right) 
\]
and determined by the imaginary part of the bath correlation function $%
\tilde{C}(t)$. Hence, the final QDT formulation is 
\begin{equation}
\dot{\rho}(t)=-i\left[ H_s-\hat{\mu}h(t)+\hat{Q}\xi (t),\rho (t)\right]
-R_s\rho (t).  \label{Liouville}
\end{equation}
The static superoperator $R_s$ is 
\[
R_s\rho (t)=\left[ Q,Q^s\rho -\rho \left( Q^s\right) ^{\dagger }\right] , 
\]
where the field-free time-independent dissipation coupling mode $Q^s$ is 
\[
Q^s=C(\hat{L}_s)Q=\frac 1{\sqrt{2}}\left( C\left( -\Omega \right)
\,a+C\left( \Omega \right) \,a^{\dagger }\right) =\frac 1{\sqrt{2}}\left(
k_{+}\,a+k_{-}\,a^{\dagger }\right) , 
\]
where $k_{+}=C\left( -\Omega \right) $ and $k_{-}=C\left( \Omega \right) $
are defined by the bath interaction spectrum $C\left( \Omega \right) $.
Using the results of Sec. III we have $C\left( \Omega \right) =J\left(
\Omega \right) n\left( \Omega \right) $ and $C\left( -\Omega \right)
=J\left( \Omega \right) \left[ n\left( \Omega \right) +1\right] $, where $%
n\left( \Omega \right) =\left( \exp \left( \beta \Omega \right) -1\right)
^{-1}$. Then the dissipation superoperator $R_s$ has the conventional
formulation 
\[
R_s\rho (t)=\frac{k_{-}}2aa^{\dagger }\rho -\frac 12\left(
k_{-}+k_{-}^{*}\right) a^{\dagger }\rho a+\frac{k_{-}^{*}}2\rho aa^{\dagger
} 
\]
\[
+\frac{k_{+}}2a^{\dagger }a\rho -\frac 12\left( k_{+}+k_{+}^{*}\right) a\rho
a^{\dagger }+\frac{k_{+}^{*}}2\rho a^{\dagger }a. 
\]
After some simple algebra, we obtain 
\[
R_s\rho (t)=J\left( \Omega \right) \left[ n\left( \Omega \right) +1\right]
\left( \frac 12\left\{ a^{\dagger }a,\rho \right\} -a\rho a^{\dagger
}\right) +J\left( \Omega \right) n\left( \Omega \right) \left( \frac 12%
\left\{ aa^{\dagger },\rho \right\} -a^{\dagger }\rho a\right) . 
\]

Using the differential representation for the Bose superoperator, one can
convert master equation (\ref{Liouville}) into Fokker-Plank equations (see the 
Appendix for details) for the Wigner function $f$

\begin{equation}
\dot{f}=\left( i\Omega +\frac \gamma 2\right) \frac \partial {\partial z}%
\left( zf\right) +\left( -i\Omega +\frac \gamma 2\right) \frac \partial {%
\partial z^{*}}\left( z^{*}f\right) +\gamma \left( n\left( \Omega \right) +%
\frac 12\right) \frac{\partial ^2f}{\partial z\partial z^{*}}-\frac{i\tilde{h%
}\left( t\right) }{\sqrt{2}}\left( \frac{\partial f}{\partial z}-\frac{%
\partial f}{\partial z^{*}}\right) ,  \label{Fokker}
\end{equation}
where $z,z^{*}$ are the complex variables, $\tilde{h}\left( t\right)
=h(t)-\xi (t)$ and $\gamma =J\left( \Omega \right) $. To derive differential
equations for Weyl symbols $\left\langle a\right\rangle _W$, $\left\langle
a\right\rangle _W^{\dagger }$ of the boson operators $a$ and $a^{\dagger }$
we multiply Eq.(\ref{Fokker}) by $z$ or $z^{*}$, respectively, and integrate
over the complex plane. Supposing that $f\rightarrow 0$ at $\left| z\right|
^2\rightarrow \infty $, we obtain the system 
\begin{equation}
\frac{\partial \left\langle a\right\rangle _W}{\partial t}=-\left( i\Omega +%
\frac \gamma 2\right) \left\langle a\right\rangle _W+\frac{i\tilde{h}(t)}{%
\sqrt{2}},  \label{sys1}
\end{equation}
\begin{equation}
\frac{\partial \left\langle a^{\dagger }\right\rangle _W}{\partial t}%
=-\left( -i\Omega +\frac \gamma 2\right) \left\langle a^{\dagger
}\right\rangle _W-\frac{i\tilde{h}(t)}{\sqrt{2}}.  \label{sys2}
\end{equation}
Using the coordinate $x=\left\langle a+a^{\dagger }\right\rangle _W/\sqrt{2}$
and the conjugated momentum $p_x=-i\left\langle a-a^{\dagger }\right\rangle
_W/\sqrt{2}$ we get the equation of the damped harmonic oscillator, 
\begin{equation}
\frac{\partial ^2x}{\partial t^2}+\gamma \frac{\partial x}{\partial t}%
+\omega _0^2x=\Omega \tilde{h}\left( t\right)  \label{Damp}
\end{equation}
with $\omega _0^2=\Omega ^2+\gamma ^2/4$, which is the dynamic equation for
the Bloch wall \cite{Neel52}.

The magnetic ac susceptibility was measured within the frequency range from
1 Hz up to 1 kHz (slow varying ac-field). In assumption that the
characteristic time scale of the experiment of order $10^{-3}\div 1$ sec is
much greater than the characteristic relaxation times $\zeta ^{-1}$ of the
bath ($\omega \ll \zeta )$, we obtain 
\[
\xi (t)\approx \frac{2\Omega \,\nu _m^{^{\prime \prime }}}{\left( \zeta
_m^2+\Omega ^2\right) \zeta _m}h(t)=\sigma h(t), 
\]
i.e. the effective local-field correction $\xi (t)$ depends on the incident
field.

Now we use Eq.(\ref{Damp}) to evaluate the range of relaxation time $\tau
\sim 1/\gamma $. We note first that without an applied bias dc-field the
oscillator frequency $\Omega $ is determined by the interchain coupling,
whereas in the bias fields of the experiment, approaching a critical value
of the AF-FI transition, $\Omega $ is reduced to much smaller values, when a
leading contribution to the ac susceptibility results from the motion of
domain walls separating AF and FI phases. To reach a consistency with the
data on a time evolution of the longitudinal magnetization in pulsed fields
(see Sec. V), we suppose $\Omega \ll \gamma $ and consider a small ac-field
frequency $\omega \ll \omega _0\sim \gamma $. Then the solution of Eq.(\ref
{Damp}) has a relaxation character that yields the expressions for $\chi
^{\prime }(\omega )$ and $\chi ^{"}(\omega )$ in the usual Debye form 
\[
\chi ^{\prime }\left( \omega \right) =\frac{(1-\sigma )\Omega }{\omega _0^2}%
\frac 1{1+\omega ^2\tau ^2}, 
\]
\begin{equation}
\chi ^{^{\prime \prime }}\left( \omega \right) =\frac{(1-\sigma )\Omega }{%
\omega _0^2}\frac{\omega \tau }{1+\omega ^2\tau ^2},  \label{dynw}
\end{equation}
$\;$where the relaxation time $\tau =\gamma /\omega _0^2$. The maximum of
the imaginary part of the ac susceptibility $\chi ^{^{\prime \prime }}\left(
\omega \right) $ is reached at $\omega _{\max }=\tau ^{-1}=\omega
_0^2/\gamma \sim \gamma $. According to the available experimental data \cite
{Bar2003} $\omega _{\max }\sim 10\div 100$ Hz (T = 3$\div $3.5 K) that
yields $\tau \sim 10\div 100$ msec for the small frequencies $\sim 1$ Hz.
This agrees qualitatively with the $\tau $ values found from the relaxation
of the remanent magnetization (Fig. 3).

\section{Saturation process: strong magnetic field pulses}

The key moment distinguishing this case from previous consideration is the
value and time-dependence of the external driving filed. The ac-field being
of order $10^{-4}T$ is weak in the sense that the system remains near global
equilibrium at all times. This is not the case for a strong field ($\sim
5\,T $) changing fast in comparison with relaxation to global equilibrium. A
long time scale of a driving field $H(t)$ ($\sim 10\,ms$) prohibits the
normal evolution towards a Boltzmann distribution of states due to dynamical
non-Markovian effects. This feature is intrinsically built into the QDT,
hence we can similarly construct a FP equation to evaluate the time
evolution of a longitudinal magnetization.

We introduce the operator $\hat{\mu}=a^{\dagger }a$ interacting with an
external pulsed field 
\[
h(t)=h_0\sin \left( \frac \pi Tt\right) \left[ \theta (t)-\theta
(t-T)\right] , 
\]
where $T$ is the time period of the external field and other definitions are
identical to that used for the ac-field. This choice ensures a saturation of
magnetization by pulsed field measurement performed in a half-pulse regime.

Unlike the ac-case the commutator contained in the integrand of Eq.(\ref
{Qt}) now becomes an operator 
\[
L_{sf}(\tau ^{\prime })G_s(\tau ^{\prime }-\tau )Q=-h(\tau ^{\prime })\left[ 
\hat{\mu},G_s(\tau ^{\prime }-\tau )Q\right] =\frac{h(\tau ^{\prime })}{%
\sqrt{2}}\left( e^{i\Omega (\tau ^{\prime }-\tau )}a-e^{-i\Omega (\tau
^{\prime }-\tau )}a^{\dagger }\right) . 
\]
Further simplicity arises from that 
\[
G(t,\tau )\,\hat{a}=\exp \left\{ i\int\limits_\tau ^t\left[ \Omega -h\left(
\tau ^{\prime }\right) \right] \,d\tau ^{\prime }\right\} \;a, 
\]
and 
\[
G(t,\tau )\,\hat{a}^{\dagger }=\exp \left\{ -i\int\limits_\tau ^t\left[
\Omega -h\left( \tau ^{\prime }\right) \right] \,d\tau ^{\prime }\right\}
\;a^{\dagger }. 
\]
Thus, we have 
\begin{equation}
G(t,\tau ^{\prime })L_{sf}(\tau ^{\prime })G_s(\tau ^{\prime }-\tau )Q=\frac{%
h(\tau ^{\prime })}{\sqrt{2}}\left[ e^{i\Omega (t-\tau )-ih_0g(t,\tau
^{\prime })}a-e^{-i\Omega (t-\tau )+ih_0g(t,\tau ^{\prime })}a^{\dagger
}\right] ,  \label{integr1}
\end{equation}
where $\int\limits_{\tau ^{\prime }}^th(\tau ^{\prime \prime })d\tau
^{\prime \prime }\equiv h_0g\left( t,\tau ^{\prime }\right) .$ Substituting (%
\ref{integr1}) into Eq.(\ref{Qt}) we obtain 
\begin{equation}
\tilde{Q}(t)=\tilde{Q}^s-\frac i{\sqrt{2}}\int\limits_0^\infty d\tau
\,\lambda (\tau )\,h(t-\tau ),  \label{Qt1}
\end{equation}
where the system-bath coupling response becomes an operator 
\[
\lambda (\tau )\equiv \int\limits_\tau ^\infty d\tau ^{\prime }\,\tilde{C}%
(\tau ^{\prime })\left[ e^{i\Omega \tau ^{\prime }-ih_0g(t,t-\tau
)}a-e^{-i\Omega \tau ^{\prime }+ih_0g(t,t-\tau )}a^{\dagger }\right] . 
\]
The explicit expression for $\lambda (\tau )$ can be easily carried out as 
\[
\lambda (\tau )=\frac{\nu _m}{\zeta _m-i\Omega }e^{i\Omega \tau -\zeta
_m\tau }e^{-ih_0g(t,t-\tau )}a-\frac{\nu _m}{\zeta _m+i\Omega }e^{-i\Omega
\tau -\zeta _m\tau }e^{ih_0g(t,t-\tau )}a^{\dagger } 
\]
via Meier-Tannor parametrization.

The convolution in Eq.(\ref{Qt1}) is simplified as 
\[
\int\limits_0^\infty d\tau \;\lambda (\tau )\,h(t-\tau )=\left\{ 
\begin{array}{c}
h_0\int\limits_{t-T}^td\tau \;\lambda (\tau )\sin \left[ \frac \pi T(t-\tau
)\right] ,\;t\geq T \\ 
h_0\int\limits_0^td\tau \;\lambda (\tau )\sin \left[ \frac \pi T(t-\tau
)\right] ,\;0\leq t<T \\ 
0,\;t<0
\end{array}
\right. 
\]
The dissipative mode is then defined as follows 
\[
\tilde{Q}(t)=\frac 1{\sqrt{2}}\left( \tilde{k}_{+}\,a+\tilde{k}%
_{-}\,a^{\dagger }\right) , 
\]
where 
\begin{equation}
\tilde{k}_{+}=k_{+}-i\frac{h_0\nu _m}{\zeta _m-i\Omega }\Phi (t),\;\tilde{k}%
_{-}=k_{-}+i\frac{h_0\nu _m}{\zeta _m+i\Omega }\Phi ^{*}(t),  \label{coefk}
\end{equation}
and 
\begin{equation}
\Phi (t)=\left\{ 
\begin{array}{c}
\int\limits_{t-T}^td\tau \;\sin \left[ \frac \pi T(t-\tau )\right]
e^{i\Omega \tau -\zeta _m\tau }e^{-ih_0g(t,t-\tau )},\;t\geq T \\ 
\int\limits_0^td\tau \;\sin \left[ \frac \pi T(t-\tau )\right] e^{i\Omega
\tau -\zeta _m\tau }e^{-ih_0g(t,t-\tau )},\;0\leq t<T \\ 
0,\;t<0
\end{array}
\right. .  \label{memory}
\end{equation}
The coefficients $k_{+}=J\left( \Omega \right) \left[ n\left( \Omega \right)
+1\right] $ and $k_{-}=J\left( \Omega \right) n\left( \Omega \right) $ are
determined as for the ac-case. The dissipation superoperator (\ref{DIS})
reads as 
\[
R(t)\rho (t)=\frac{\tilde{k}_{-}}2aa^{\dagger }\rho -\frac 12\left( \tilde{k}%
_{-}+\tilde{k}_{-}^{*}\right) a^{\dagger }\rho a+\frac{\tilde{k}_{-}^{*}}2%
\rho aa^{\dagger } 
\]
\[
+\frac{\tilde{k}_{+}}2a^{\dagger }\hat{a}a-\frac 12\left( \tilde{k}_{+}+%
\tilde{k}_{+}^{*}\right) a\rho a^{\dagger }+\frac{\tilde{k}_{+}^{*}}2\rho
a^{\dagger }a. 
\]
The Liouville equation 
\[
\dot{\rho}(t)=-i\left[ H_s-\hat{\mu}h(t),\rho (t)\right] -R\rho (t) 
\]
takes the final form 
\[
\dot{\rho}(t)-i\tilde{\Omega}\left[ \rho ,a^{\dagger }a\right] =-\frac{%
\tilde{k}_{-}}2aa^{\dagger }\rho +\frac 12\left( \tilde{k}_{-}+\tilde{k}%
_{-}^{*}\right) a^{\dagger }\rho a-\frac{\tilde{k}_{-}^{*}}2\rho aa^{\dagger
} 
\]
\begin{equation}
-\frac{\tilde{k}_{+}}2a^{\dagger }a\rho +\frac 12\left( \tilde{k}_{+}+\tilde{%
k}_{+}^{*}\right) a\rho a^{\dagger }-\frac{\tilde{k}_{+}^{*}}2\rho
a^{\dagger }a,  \label{MaEq1}
\end{equation}
where $\tilde{\Omega}=\Omega -h(t)$.

Then we convert Eq.(\ref{MaEq1}) into the equivalent Fokker-Plank equation
using the Wigner functions for the density matrix $\rho $ and Bose operators 
\begin{equation}
\frac{\partial \bar{n}}{\partial t}=-\frac 1{\sqrt{2}}\left( \tilde{k}_{+}+%
\tilde{k}_{+}^{*}-\tilde{k}_{-}-\tilde{k}_{-}^{*}\right) \left( \bar{n}+%
\frac 12\right) +\frac 1{2\sqrt{2}}\left( \tilde{k}_{+}+\tilde{k}_{+}^{*}+%
\tilde{k}_{-}+\tilde{k}_{-}^{*}\right) ,  \label{FP2}
\end{equation}
where $\bar{n}=\left\langle a^{\dagger }a\right\rangle _t$ is the number of
oscillator excitations corresponding to the instant magnetization. In the
complete absence of the external field ($h_0=0$) Eq.(\ref{FP2}) amounts to 
\[
\frac{\partial \bar{n}}{\partial t}=-\frac 1{T_1}\left( \bar{n}-n_0\right) , 
\]
and the net magnetization relaxes from an initial value to the equilibrium
one $n_0=n(\Omega )=\left( e^{\beta \Omega }-1\right) ^{-1}$ with the
spin-lattice relaxation rate $T_1^{-1}=\sqrt{2}\gamma $. In general, by
using the coefficients (\ref{coefk}), we can recast Eq.(\ref{FP2}) as 
\begin{equation}
\frac{\partial \bar{n}}{\partial t}=-\frac 1{T_1}\left( \bar{n}-n_0\right) -%
\sqrt{2}\bar{n}\,f_1(t)+f_2(t)-f_1(t),  \label{FPrel}
\end{equation}
where the time-dependent coefficients are 
\[
f_1(t)=\frac{\sqrt{2}h_0\nu _m^{^{\prime \prime }}}{\zeta _m^2+\Omega ^2}%
\left( \zeta _m\Phi ^{^{\prime }}(t)-\Omega \Phi ^{^{\prime \prime
}}(t)\right) ,\;f_2(t)=\frac{\sqrt{2}h_0\nu _m^{\prime }}{\zeta _m^2+\Omega
^2}\left( \Omega \Phi ^{^{\prime }}(t)+\zeta _m\Phi ^{^{\prime \prime
}}(t)\right) . 
\]
As can be inferred from Eq.(\ref{memory}), $f_{1,2}(t)$ decreases
exponentially with time and falls to zero as $t>\zeta ^{-1}$.

The differential equation (\ref{FPrel}) can be solved numerically with the
initial condition $\bar{n}(t_0)=0$ ($t_0<0$). The results for the simplest
one-exponential case $m=1$ are presented in Fig. 4 where the experimental
data are plotted for comparison. We can see how the magnetization, following
the $h(t)$ variation, increases initially with time. After switching off the
field, the magnetic moment of the system has a two-step evolution. The
first, rapid stage ends when the system arrives at the critical state due to
the balance of the magnetic driving force and the coercive force on the
domain walls. The second, slow stage of the evolution is due to backward
domain wall movement accompanied with damped oscillations around the moving
center position. The damping indicates the effect of spin-lattice coupling.
At rather low temperatures below 3D ordering, most of the system energy is
lost to the bath because of dissipation. It is obviously seen from the inset
of Fig.4 that the damping is governed by the force fluctuation decay $\zeta $
in the bath. The slower the decay rate the more prominent domain wall
oscillations.

\section{Conclusions}

In conclusion, we summarize briefly the results presented in the paper.
Measurements of the magnetization in pulsed fields for the
molecule-based magnetic complex [Mn(hfac)$_{\text{2}}$BNO$_{\text{H}}$]
composed from antiferromagnetically ordered ferrimagnetic chains show that
the change in the magnetization in this compound below 3D ordering temperature
is a slow dynamical process controlled by motion of magnetic domains with a
single energy barrier height. As a critical field for the direct AF-FI
transition is approached during application of a pulsed field, the domain
wall motion along the separate field unfavorable chains starts to develop.
After the field is switched off, the inverse transition from the
field-induced FI state to the initial AF state in zero field 
provides very slow dynamics of the magnetization. The energy barrier
hampering the reversal of the magnetization is originated from the
antiferromagnetic interchain coupling. The latter thus plays an analogous
role to that of magnetic anisotropy in a molecular cluster or a single chain
magnet. The measurements of the ac-susceptibility in the region of the
metamagnetic transition, where the AF and FI\ phase coexist, show that a
leading contribution to the ac susceptibility results from the motion of
domain walls separating the AF and FI phases.

The domain-wall motion in both the ac and pulsed fields is
accompanied by energy losses that causes a local heating of the samples.
This is because the energy of the magnetic system transforms into the phonon
energy, and, as a consequence, 3D magnetic ordering holds. Thus the
system-bath coupling is a crucial in the description of the relaxation
dynamics in [Mn(hfac)$_{\text{2}}$BNO$_{\text{H}}$].

On the basis of quantum dissipative theory in the standard second-order approximation 
for the system-bath Hamiltonian, we derive Fokker-Plank equations
for observables of the reference system. It is known that this well-justified 
approximation makes applicable the resultant FP equation in a large number of 
dissipative systems  provided the system-bath coupling is not strong. The QDT-FP 
formalism has advantages of application convenience and straightforwardness, 
as well as the ability to address both saturation processes caused by strong 
magnetic field pulses and processes far from a saturation by a small 
oscillating ac-field.

The complex magnetic ac susceptibility is calculated within an approach,
where the magnetization is controlled by the field-induced sideways motion
of domain walls. The expressions for $\chi ^{\prime }(\omega )$ and $\chi
^{\prime \prime }(\omega )$ have the usual Debye form for small frequencies. 
>From the maximum of the imaginary part of the ac susceptibility we evaluate the relaxation time
that agrees qualitatively with that found from the remanent magnetization
data.

In the case of a small oscillating field the system remains near global
equilibrium at all times, whereas a strong long-time driving field changing
fast in comparison with relaxation to global equilibrium prohibits the
normal evolution towards a Boltzmann distribution due to dynamical
non-Markovian effects. In order to obtain a reliable understanding of the
physics of the process we derive a FP\ equation in the framework QDT. The
study of a time relaxation of a longitudinal magnetization shows that it
experiences a two-step evolution after the field is switched off. The first
rapid stage ends when the system arrives at the critical state where the
magnetic driving force and the coercetive force acting jointly on the domain
wall are balanced. The second slow stage of the evolution corresponds to
backward domain wall movement together with damped oscillations around the
moving domain-wall center. The damping is managed by a decay of force-force
correlations of Langevin dissipative modes acting on the system from the
bath.

\acknowledgments
We would like to thanks E.Z. Kuchinskii for discussions. This work was
partly supported by the grant NREC-005 of US CRDF (Civilian Research \&
Development Foundation). Two of us (V.E.S. and A.S.B.) thanks the Foundation
Dynasty (Moscow) for the support.

\section{Appendix A.}

As a method for expressing the density operator in terms of c-number
functions, the Wigner functions often lead to considerable simplification of
the quantum equations of motion, as for example, by transforming operator
master equations into more amenable Fokker-Plank differential equations. By
the Wigner function one can express quantum-mechanical expectation values in
form of averages over the complex plane (the classical phase space), the
Wigner function playing the role of a c-number quasi-probability
distribution 
\[
f(z,z^{*};t)=Tr\left[ \rho (t)\hat{\delta}_W\left( a-z\right) \right] , 
\]
where 
\[
\hat{\delta}_W\left( a-z\right) =\int \frac{d^2x}{\pi ^2}\exp \left\{
ix^{*}\left( a-z\right) +ix\left( a^{\dagger }-z^{*}\right) \right\} 
\]
is the operator delta function with $d^2x=dx_1dx_2$, $x=x_1+ix_2,$ $%
z=z_1+iz_2$. The Wigner function has the following property $\int
f(z,z^{*};t)\,d^2z=1$ and allows to easily evaluate expectations of
symmetrically ordered products of the field operators, corresponding to
Weyl's quantization procedure \cite{Perel}

\[
\int \left( z^{*}\right) ^mz^nf(z,z^{*};t)\,d^2z=\left\langle \left\{ \left(
a^{\dagger }\right) ^ma^n\right\} \right\rangle , 
\]
where 
\[
\left\{ (a^{\dagger })^ma^n\right\} =\frac 1{\left( m+n\right) !}%
\sum\limits_PP(a^{\dagger })^ma^n 
\]
and the symbol $P$ denotes a permutation of the Bose operators.

Using the last identity one obtains 
\[
\int z^{*}f(z,z^{*};t)\,d^2z=\left\langle a^{\dagger }\right\rangle
_t,\;\int z\,f(z,z^{*};t)\,d^2z=\left\langle a\right\rangle _t, 
\]
and 
\[
\int z^{*}zf(z,z^{*};t)\,d^2z=\left\langle a^{\dagger }a\right\rangle
_t+1/2. 
\]
The Weyl symbol for any operator $\hat{O}$ is determined by 
\[
\left( \hat{O}\right) _W(z,z^{*})=\pi Tr\left[ \hat{O}\hat{\delta}_W\left( 
\hat{a}-z\right) \right] , 
\]
and the inversion formula is

\[
\hat{O}=\int \left( \hat{O}\right) _W(z,z^{*})\hat{\delta}_W\left( \hat{a}%
-z\right) d^2z. 
\]

The Wigner functions of multiplication of two operators $\hat{A}\hat{B}$ can
be easily obtained from those of $\left( \hat{A}\right) _W$ and $\left( \hat{%
B}\right) _W$ using the following identities 
\[
\left( \hat{A}\hat{B}\right) _W\left( z,z^{*}\right) =\left( \hat{A}\right)
_W\left( z+\frac 12\frac \partial {\partial u^{*}},z^{*}-\frac 12\frac 
\partial {\partial u}\right) \left( \hat{B}\right) _W\left. \left(
u,u^{*}\right) \right| _{u=z}, 
\]
\[
\left( \hat{A}\hat{B}\right) _W\left( z,z^{*}\right) =\left( \hat{B}\right)
_W\left( z-\frac 12\frac \partial {\partial u^{*}},z^{*}+\frac 12\frac 
\partial {\partial u}\right) \left( \hat{A}\right) _W\left. \left(
u,u^{*}\right) \right| _{u=z} 
\]
and Weyl symbols for Bose operators 
\[
\left( a^{\dagger }\right) _W\left( z,z^{*}\right) =z^{*},\;\left( a\right)
_W\left( z,z^{*}\right) =z. 
\]
The relation between the Wigner function and the Weyl symbol of the density
operator is the following 
\[
f(z,z^{*};t)=\pi ^{-1}\left( \rho _S(t)\right) _W\left( z,z^{*}\right) . 
\]

\newpage

\begin{figure}[tbp]
\caption{Time dependence of the magnetization of [Mn(hfac)$_{\text{2}}$BNO$_{%
\text{Cl}}$] (a) and [Mn(hfac)$_{\text{2}}$BNO$_{\text{H}}$] (b) compounds
at 1.5 K after application of the 5 T pulsed field. The logarithmic curve in
the first case results from a relaxation mechanism that involves a
distribution of energy barriers. The exponential decrease for the second
case arises from a single-barrier activation mechanism.}
\label{fig1}
\end{figure}

\begin{figure}[tbp]
\caption{In an external positive field the system consists of field-favorable
(white) and field-unfavorable (shaded) chains. Interface kinks (domain
walls) in the second-type chains can move only along the dotted arrows.}
\label{fig2}
\end{figure}

\begin{figure}[tbp]
\caption{Temperature dependence of the relaxation time obtained from the
remanent magnetization $M_r$ at the value $M_r=1$ A m$^2$/kg after
application of the 4 T pulsed field. The solid line shows Arrhenius behavior
at lower temperatures.}
\label{fig3}
\end{figure}

\begin{figure}[tbp]
\caption{Model time dependence of the magnetization during and after
application of the pulsed field found from the Fokker-Plank equation (dotted
line). The experimental magnetization curve for [Mn(hfac)$_{\text{2}}$BNO$_{%
\text{H}}$] at T=1.61 K presented for comparison is shown as a solid line.
The profile of the pulsed field used in the measurement is also plotted. The
model parameters are $\nu _m^{\prime }=0.3$, $\nu _m^{^{\prime \prime
}}=0.95 $, $T=12$, $\gamma =0.004$, $\Omega =1.3$, $\zeta _m=0.7$, $h_0=1.3$%
. Inset: time dependence of the longitudinal magnetization when $\gamma
=0.003$, $\zeta _m=0.4$, $\nu _m^{^{\prime \prime }}=0.35$. The damped
domain wall oscillations are resolved more clearly.}
\label{fig4}
\end{figure}


\begin{references}
\bibitem{Snoska}  It goes without saying, the field strength used, up to 5 T
in the ac-susceptibility measurements and up to 42 T for magnetization
measurements in pulsed fields, is not enough to destroy the antiparallel
alignment between spins of the Mn and the NO group and to induce
ferromagnetic ordering in the chains.
\end{references}
\end{document}